\documentclass{pasj01}
\draft 
\Received{$\langle$20-Jan-2016$\rangle$}
\Accepted{$\langle$03-Mar-2016$\rangle$}
\Published{$\langle$publication date$\rangle$}
\begin{document}
\title{Galactic Center Mini-spiral  by ALMA - Possible Origin of the Central Cluster}
\author{Masato Tsuboi$^{1, 3}$, Yoshimi Kitamura$^1$, Makoto Miyoshi$^2$, Kenta Uehara$^3$,  Takahiro Tsutsumi$^4$ and Atsushi Miyazaki$^{2, 5}$}%
\altaffiltext{1}{Institute of Space and Astronautical Science, Japan Aerospace Exploration Agency,\\
3-1-1 Yoshinodai, Chuo-ku, Sagamihara, Kanagawa 252-5210, Japan }
\email{tsuboi@vsop.isas.jaxa.jp}
\altaffiltext{2}{National Astronomical Observatory of Japan, Mitaka, Tokyo 181-8588, Japan}
\altaffiltext{3}{Department of Astronomy, The University of Tokyo, Bunkyo, Tokyo 113-0033, Japan}
\altaffiltext{4}{National Radio Astronomy Observatory,  Socorro, NM 87801-0387, USA}
\altaffiltext{5}{Japan Space Forum, Kandasurugadai, Chiyoda-ku,Tokyo,101-0062, Japan}

\maketitle

\begin{abstract}
We present  continuum images of the ``Galactic Center Mini-spiral" of 100, 250, and 340 GHz bands with the analysis of the Cy.0 data  acquired from the Atacama Large Millimeter/submillimeter Array (ALMA) archive. Pretty good UV coverage of the data and the ``self-calibration" method give us an opportunity to obtain dynamic ranges of over $2\times10^4$ in the resultant maps of the 250 and 340 GHz bands. In particular the image of the 340 GHz band has high dynamic ranges unprecedented in sub-millimeter wave.  The angular resolutions attain to $1.57\arcsec\times1.33\arcsec$ in the 100 GHz band, $0.63\arcsec\times0.53\arcsec$  in the 250 GHz band, and $0.44\arcsec\times0.38\arcsec$ in the 340 GHz band, respectively.  
The continuum images clearly depict the ``Mini-spiral", which is an ionized gas stream in the vicinity of Sgr A$^{\ast}$. We found the tight correlation between the dust emission peaks and the OB/WR stars in the Northern-arm of the ``Mini-spiral". The core mass function of the dust core identified by the {\it clumpfind} algorithm would obey the flat power-law $dN/dM\propto M^{-1.5\pm0.4}$ on the high-mass side.
These support the scenario that the star forming cloud has fallen into the immediate vicinity of Sgr A$^{\ast}$ for the origin of the Central cluster.

\end{abstract}

\section{Introduction}
The Galactic center region is the nucleus of the nearest spiral galaxy; the Milky Way. Sagittarius A$^{\ast}$ (Sgr A$^{\ast}$) is a compact source from radio to X-ray located near the dynamical center of the galaxy and is associated with the Galactic central supermassive black hole (GCBH). The mass of the GCBH is estimated to be $M_{\mathrm{GC}}\simeq 4\times10^6M_\odot$  from infrared (IR) astrometry observations (e.g. \cite{Ghez}, \cite{Gillessen}). The region surrounding Sgr A$^{\ast}$ is recognized to be a laboratory for peculiar phenomena, which will be found in the nuclei of normal galaxies by future telescopes. 
In the last two decades, young and highly luminous clusters have been found in the Central Molecular Zone (CMZ) (\cite{MorrisSerabyn}) by IR observations  (e.g.  \cite{Genzel}, \cite{Figer1999}, \cite{Figer2002}).  These bright star clusters except for the Central cluster presumably formed in the cradle dense molecular clouds in the CMZ. 
However, the Central cluster is a unique object in the Milky Way because the cluster is centered at Sgr A$^{\ast}$ and concentrates within $r\sim 0.5$pc. The cluster contains $\sim100$ OB and WR stars  (e.g. \cite{Genzel}).   It may be difficult to make the Central cluster in a way by which stars are usually formed in the disk region because of the following reasons.
First, the tidal force of Sgr A$^{\ast}$ must have a serious effect on the star formation since the minimum number H$_2$ density for stabilization toward the tidal  shearing  is $n(\mathrm{H_2})\gtrsim3\times10^8$ cm$^{-3}$ at $r\sim 0.5$pc (\cite{Christopher}, \cite{Tsuboi2011}). Second, the strong Lyman continuum radiation from the early type stars in the cluster ionizes rapidly the ISM in the region. The ionized ISM  stream is identified as the ``Galactic Center Mini-spiral" surrounding Sgr A$^{\ast}$(\cite{Ekers1983}, \cite{LO1983}), of which elongated appearance and kinematics indicate that it is  a tentative structure. 
Therefore, it is still an open question how the Central cluster has formed.

Two distinct scenarios for the formation of the Central cluster have been proposed so far.  One is current in-situ star formation in such extreme environment of the vicinity of Sgr A$^{\ast}$.  The other is that a molecular cloud is fallen from the region somewhat far from Sgr A$^{\ast}$ to the vicinity of Sgr A$^{\ast}$ after star formation in the cloud started (e.g. \cite{Tsuboi2015b}).
Recently a high-resolution observation ($0.088\arcsec\times0.046\arcsec$) within $30\arcsec$ of  Sgr A$^{\ast}$ at 34 GHz with JVLA detected many sources with  bow-shock appearance (\cite{Yusef-Zadeh2015a}).  The authors explained them as ionized outer envelopes of newly forming low mass stars. If so, this may be circumstantial evidence of the in-situ star formation near Sgr A$^{\ast}$.  However, the elapsed time for falling from $r\sim 10$ pc to the vicinity of Sgr A$^{\ast}$ in the second scenario is  $10^5$ yr, comparable to the ages of protostars in the Galactic disk region. 
The low mass stars may remain as protostars if the low mass star formation in the falling cloud is as slow as that in the Galactic disk region. 
Consequently, the JVLA-detected sources do not necessarily support the first scenario of the current in-situ star formation near Sgr A$^{\ast}$ (also see \cite{Paumard}).

The detailed observations of  the survived molecular gas or partially ionized gas are necessary to understand the star formation in the the vicinity of Sgr A$^{\ast}$.  Recently SiO(5-4) emission line observations detected several spots of molecular gas around  Sgr A$^{\ast}$. However the association between the spots and the IR-detected stars are not clear (\cite{Yusef-Zadeh2013}, \cite{Yusef-Zadeh2015b}). 
We analyze the ALMA Cy.0 data of the region to depict the dust emission from the molecular cloud in the high resolution less than $1\arcsec$. Using them, we can trace the molecular cloud before destructed completely by the UV continuum radiation. 
Throughout this paper, we adopt 8.5 kpc as the distance to the Galactic center. Then, $1\arcsec$ corresponds to about 0.04 pc at the distance. 

\section{Data Analysis and Results}
We analyze the Cy.0 data of the Sgr A$^{\ast}$ region in 100-, 250-, and 340-GHz bands, which was acquired from the ALMA archive (ADS/JAO.ALMA\#2011.0.00887.S).   Nineteen 12-m antennas were available. In addition, the epoch, May 18, 2012, was suitable for imaging observation because Sgr A$^{\ast}$ was in a relatively quiet phase (\cite{Brinkerink}). Since the original observation aimed for the measurement of the time variation of Sgr A$^{\ast}$, the time span of the observation is longer than 7 hrs (from 03:30:47 UT to 10:52:16 UT). Each frequency band has four spectral windows.  The band width of the spectral window is 1.875 GHz.  
The  fields of view (FOVs) are centered at Sgr A$^{\ast}$, $\alpha_{\rm J2000}=17^h45^m40^s.04$, $\delta_{\rm J2000}=-29^\circ 00\arcmin28\arcsec.10$. NRAO530 and J1924-292 were used as phase calibrators. The flux density scale was determined using Titan and Neptune. 

Further data calibration and imaging of the archival data were done by classic VLBA AIPS(NRAO) using the “self-calibration” method.  First we removed residual fringe rates and delays of the raw visibility by the task FRING, which was found to be quite effective.
Second the residual complex gain errors of the data were minimized using the ``self-calibration" method in the software. The good UV coverage and ``self-calibration" method give us an opportunity to obtain high dynamic range  in the resultant maps.  
\begin{figure}
 \begin{center}
  \includegraphics[width=8cm]{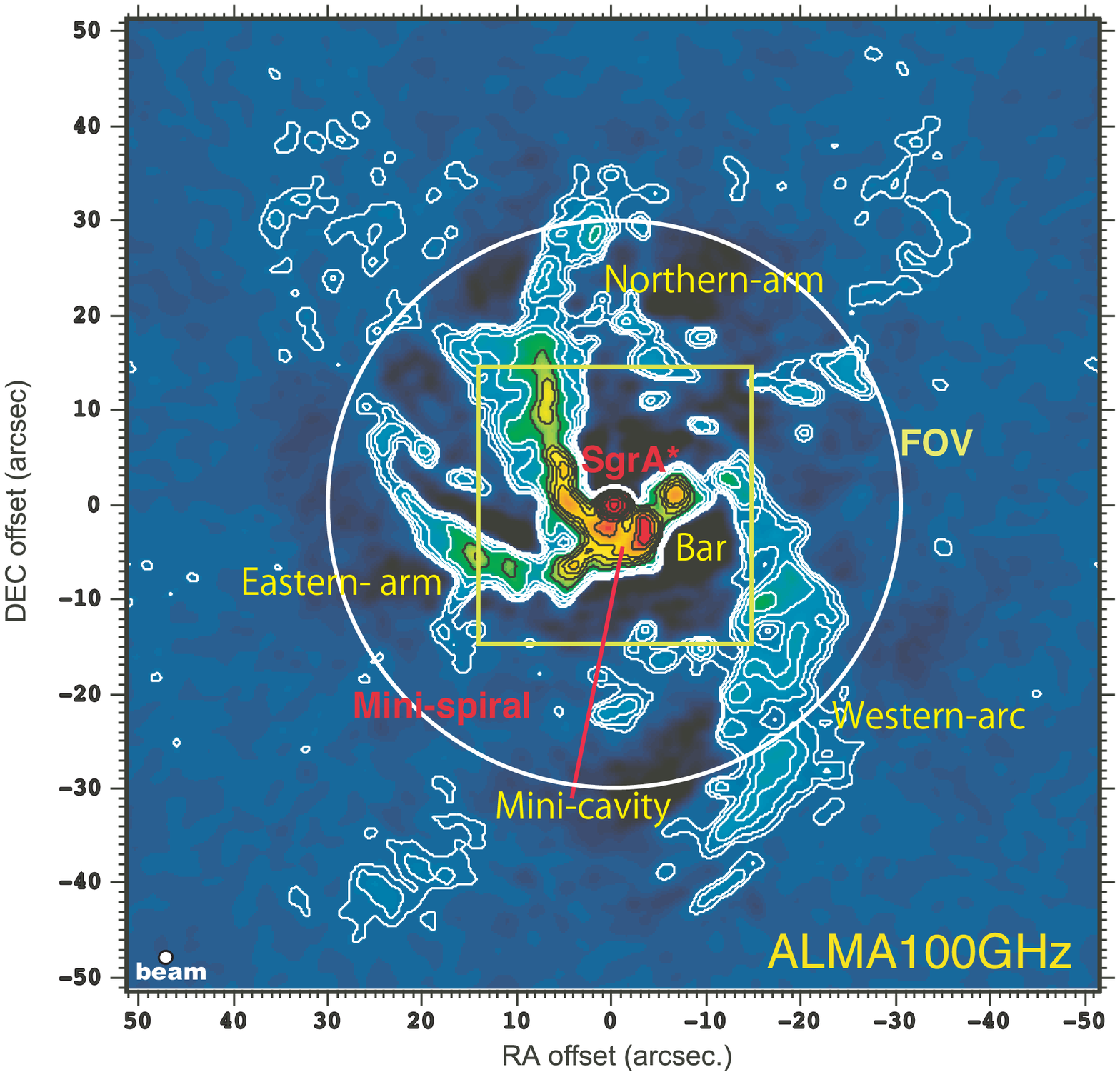}
 \end{center}
 \caption{ALMA map in the 100 GHz band of the ``Mini-spiral" including Sgr A$^{\ast}$. The four spectral windows of $f_c=93, 95, 105,$ and $107$ GHz are combined to improve the sensitivity. The diameter of the FOV is $60\arcsec$.  The angular resolution is $1.57\arcsec\times1.33\arcsec$ at $PA=74^\circ$, which is shown as an oval on the left bottom corner. The rms noise level is 0.33 mJy beam$^{-1}$, and the contour levels are 0.63, 1.3, 2.5, 5.0, 7.5, 10, 20, 30, 40, 50, 100, 200, 300, 400, 500, 1000, and 2000 mJy beam$^{-1}$. The flux density of Sgr A$^{\ast}$ is $S_\nu=2.35\pm0.19$ Jy  at 100 GHz. The square shows the area covered by Fig.2.}\label{Fig1}
\end{figure}
\begin{figure}
 \begin{center}
  \includegraphics[width=16cm]{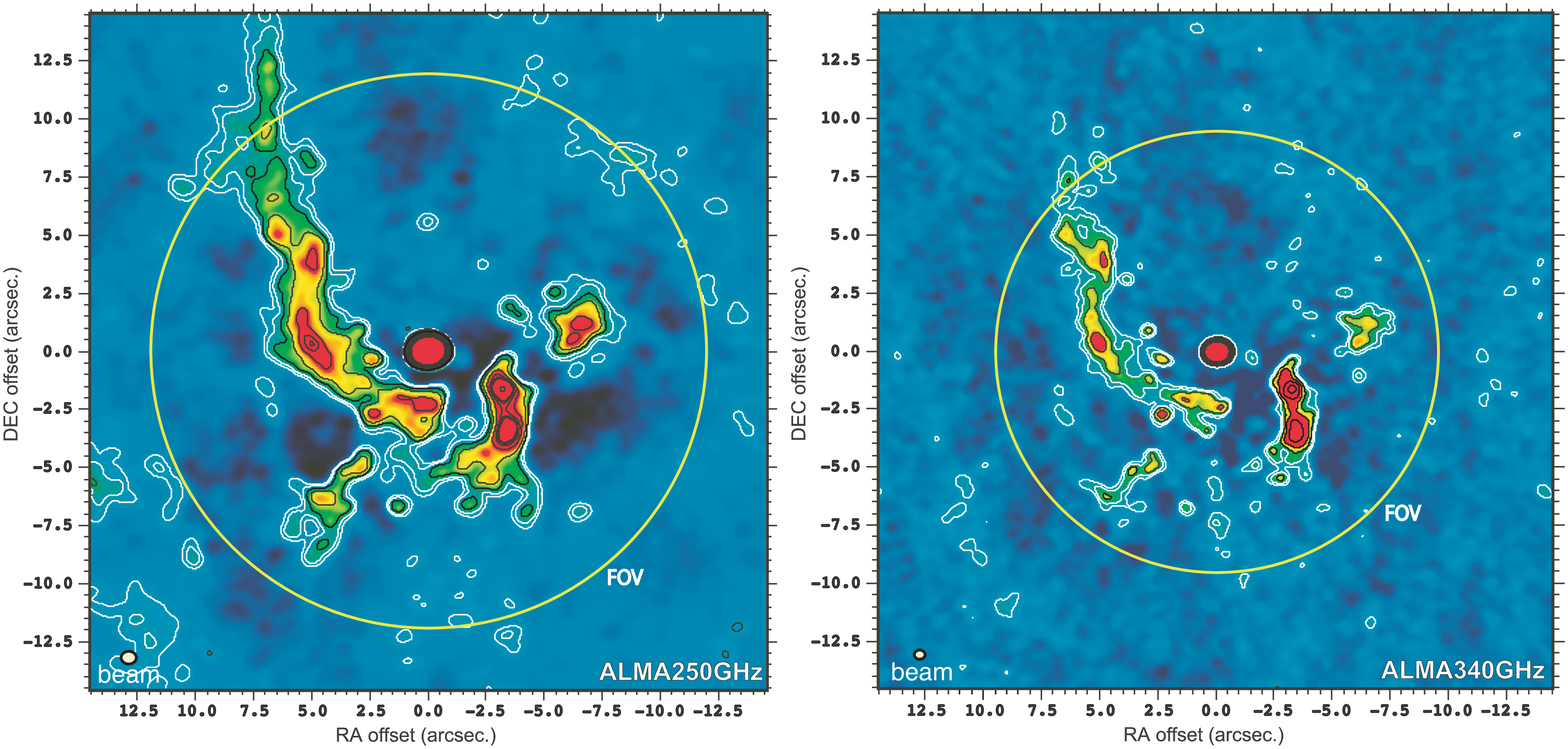}
  \end{center}
\caption{{\bf left panel}:  ALMA map in the 250 GHz band of the ``Mini-spiral"  including Sgr A$^{\ast}$. The  four spectral windows of $f_c=245, 247, 257,$ and $259$ GHz are combined to improve the sensitivity. The diameter of the FOV  is $24\arcsec$ (circle). The angular resolution is $0.63\arcsec\times0.53\arcsec$ at $PA=-84^\circ$, which is shown as an oval on the  left bottom corner. The rms noise level is 0.13 mJy beam$^{-1}$, and the contour levels are 0.31, 0.63, 1.3, 2.5, 5.0, 10, 20, 30, 40, 50, and 75 mJy beam$^{-1}$. The flux density of Sgr A$^{\ast}$ is $S_\nu=3.55\pm0.35$ Jy at 250 GHz. 
{\bf right panel}:  ALMA map in the 340 GHz band of the same region as the left panel. The  four spectral windows of $f_c=336, 338, 348,$ and $350$ GHz are combined to improve the sensitivity. The diameter of the FOV is $18\arcsec$ (circle). The angular resolution is $0.44\arcsec\times0.38\arcsec$ at $PA=-89^\circ$, which is shown as an oval on the left bottom corner.  The rms noise level is 0.33 mJy beam$^{-1}$, and the contour levels are the same as in the left panel. The flux density of Sgr A$^{\ast}$ is $S_\nu=3.44\pm0.51$ Jy at 340 GHz.}
\label{Fig2}
\end{figure}

Fig. 1 shows the intensity map of the 100 GHz band. Here, we combined the four spectral window maps of $f_c=93, 95, 105,$ and $107$ GHz in the 100 GHz band to improve the sensitivity.
 The figure size is $100\arcsec\times100\arcsec$. The resultant Full Width at Half Maximum (FWHM) beam is $1.57\arcsec\times1.33\arcsec (PA=74^\circ)$, which corresponds to $0.065{\rm pc}\times0.055{\rm pc}$ at the Galactic center. The diameter of the FOV is $D_{\rm FOV}=60\arcsec$ or 2.5 pc.  This means that the sensitivity of the map decreases to 50\% at the edge of the FOV, which is caused by the beam pattern of the element antenna. The correction for the primary beam attenuation is not applied to the map. Features with spatial scales larger than $38\arcsec $ or 1.6 pc were resolved out.  The rms noise level is $S_\nu =0.33$ mJy beam$^{-1}$ (4ch-combined) at the map center. 
The flux density of Sgr A$^{\ast}$ is $S_\nu=2.35\pm0.19$ Jy at 100 GHz. The error includes the calibration error.
The ``Mini-spiral" is clearly seen in this map, which has  the ``Eastern- arm", the ``Northern-arm", the ``Bar", and the ``Western-arc" as substructures. Moreover, the ``Mini-cavity" is identified on the Bar. Because the continuum emission at 100 GHz is originated  mainly from the ionized ISM, the appearance closely resembles that at centimeter wave (e.g. \cite{Yusef-Zadeh1987}). 

The left and right panels of Fig.2 show the intensity maps of the 250 and 340 GHz bands, respectively. Both figures cover a $29\arcsec\times29\arcsec$ area centered at Sgr A$^{\ast}$.
For the left panel, there are four spectral windows of $f_c=245, 247, 257,$ and $259$  GHz in the 250 GHz band and the four data sets are combined to improve the sensitivity. The diameters of the FOV is $D_{\rm FOV}=24\arcsec$ or 1 pc. The correction for the attenuation of the primary beam is not applied to the map. The resultant FWHM beam is $0.63\arcsec\times0.53\arcsec (PA=-84^\circ)$, or $0.026{\rm pc}\times0.022{\rm pc}$  at the Galactic center. Features with spatial scales larger than $15\arcsec $, or 0.63 pc,  were resolved out. The rms noise level at the map center is $S_\nu =0.13$ mJy beam$^{-1}$ at 250 GHz.  
The good UV coverage and ``self-calibration" give us to obtain high dynamic range of  $2.7\times10^4$ in the resultant map. The flux density of Sgr A$^{\ast}$ is $S_\nu=3.55\pm0.35$ Jy at 250 GHz. Because the light curves at 250 and 340 GHz show that Sgr A* has considerable variability, the quoted values are considered as the time-averaged flux densities at this epoch.  The error includes the calibration error.  

For the right panel, there are four spectral windows of $f_c=336, 338, 348,$ and $350$  GHz in the 340 GHz band and the four data sets are also combined to improve the sensitivity. The diameters of the FOV is $D_{\rm FOV}=18\arcsec$, or 0.75 pc. The correction for the attenuation of the primary beam is not applied to the map. The resultant FWHM beam  at 340 GHz is $0.44\arcsec\times0.38\arcsec (PA=-89^\circ)$, which corresponds to  $0.018{\rm pc}\times0.016{\rm pc}$.  Features with spatial scales larger than $11\arcsec $, or 0.46 pc, were resolved out. The rms noise level at the map center is $S_\nu =0.15$ mJy beam$^{-1}$ at 340 GHz. The dynamic range in the resultant map reaches to $2.3\times10^4$, which is unprecedented at that frequency. The flux density of Sgr A$^{\ast}$ is  $S_\nu=3.44\pm0.51$ Jy at 340 GHz. 

The Northern-arm, the western half of the Eastern-arm, and the Bar are clearly seen both in the figures at 250 and 340 GHz.  The Mini-cavity is clearly identified around  $\Delta\alpha=-1.5\arcsec$, $\Delta\delta=-2.0\arcsec$ (Cf. Fig. 6 in \cite{Zhao}). While the Eastern-half of the Eastern-arm is fairly in the FOV  at 250 GHz, it is out of the FOV  at 340 GHz. The Western-arc is out of both the FOVs. 
The Northern-arm seems to have a helical appearance in the both figures (Cf.  Fig. 22 in \cite{Zhao}). Moreover, we find many smaller components both in the figures, which have statistical significance of $>5\sigma$. Because the data in the two bands have different UV coverage, these small objects are real substructures rather than artifacts caused by the side lobes. However, these are not associated with the SiO sources detected by ALMA (\cite{Yusef-Zadeh2013}, \cite{Yusef-Zadeh2015b}). 

\begin{figure}
 \begin{center}
  \includegraphics[width=15cm]{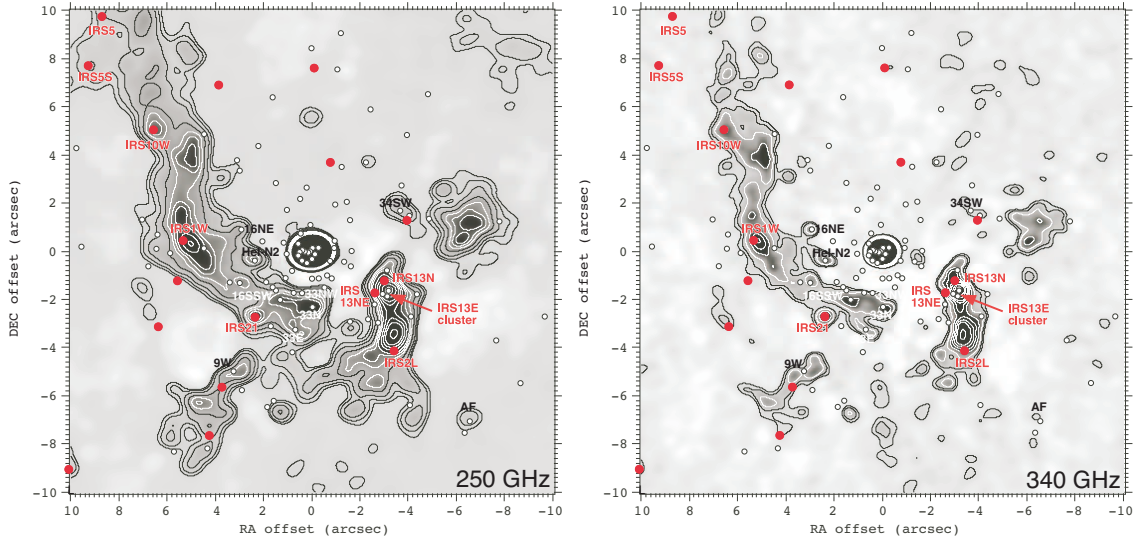}
  \end{center}	
\caption{ Positions of the early type stars detected by IR observations are plotted by  the circles on the ALMA maps at 250 GHz (left) and 340 GHz (right).  IRS1W, IRS2L, IRS5, IRS10W, IRS13EN, IRS13N, and IRS21, which are known as "bow-shock" stars designated by the filled circles. They are located on the dust emission peaks or within $0.5\arcsec$ of them in the Northern-arm and the Bar. The other early type stars are designated by the open circles. IRS9W, IRS16NE, IRS16SSW, IRS33E, IRS33N, IRS33NW,  IRS34SW, HeI-N2, and AF are also located on the dust emission peaks or within $0.5\arcsec$ of them.  }
\label{Fig3}
\end{figure}
\section{Discussion}
The continuum emission at 340 GHz can be considered as the dust thermal emission for the following reason.
The widths of the  Northern-arm at 250 and 340 GHz seem to be narrower than that at 100 GHz. In the later, the Northern-arm has a ``spine" structure and a relatively smooth envelope structure around it.  While the Northern-arm  at 250 and 340 GHz has only a ``spine" structure structure (see Fig.1 and Fig. 2). 
The beam-deconvolved widths at $\Delta\delta=2.7-3.9\arcsec$ are $\theta=\sqrt{(Gaussian~fit~width)^2-b_{maj.}\times b_{min.}}=1.30\pm0.03\arcsec$ at 250 GHz and $\theta=1.15\pm0.05\arcsec$ at 340 GHz although the width is $\theta=1.86\pm0.02\arcsec$ at 100 GHz. The beam sizes at the three bands are much smaller than the arm widths. 
In addition, the resolve-out limits at the three bands are much larger than the arm widths. Then the difference among the widths is not caused by the beam size difference and/or the resolve-out limit difference.  
While the continuum emission  at 100 GHz of the Mini-spiral is  thought to mainly come from ionized gas, the continuum emission at 340 GHz presumably comes from dust although that at 250 GHz may be a mixture of free-free emission from ionized gas and dust emission. The observed ``spine" structure  presumably shows the dust ridge in the Mini-spiral. On the other hand, the envelope structure probably shows the distribution of the ionized gas oozing from the ridge by strong UV radiation. 

To explore the relation between the dust ridge of the Mini-spiral and  the OB/WR stars of the Central cluster (e.g. \cite{Genzel}, \cite{Paumard}, \cite{Zhao}), we show the positions of the stars on the emission maps at 250 and 340 GHz in the left and right panels of Fig.3, respectively.
More than 10 early type stars are located on the dust emission peaks in the Northern-arm and the Bar.   
The ``bow-shock" stars, which have shell-like envelopes, were found in recent IR observations (e.g. \cite{Sanchez-Bermudez}).  IRS1W, IRS2L, IRS5, IRS10W, IRS13EN, IRS13N, and IRS21 seem to lie along the Northern-arm and the Bar. Moreover, they  are located on the dust emission peaks or within $0.5\arcsec$ of them except for IRS 5, which is out of the FOV at 340 GHz. Other early type stars, IRS16NE, IRS16SSW, IRS33E, IRS33N, IRS33NW,  IRS34SW, HeI-N2,  and AF, are  also  located on the dust emission peaks or within $0.5\arcsec$ of them. 
In addition, the ``famous" massive dense star  cluster, IRS13E,    is located on the center of the brightest dust condensation in the Bar (e.g. \cite{Schodel}). 
There seems to be tight correlation between the dust emission peaks and the OB/WR stars in the Northern-arm and the Bar.
On the other hand, the correlation between the Eastern-arm and the OB stars except for IRS9W  is not clear because of the limitation of the FOV.

\begin{figure}
 \begin{center}
  \includegraphics[width=9cm]{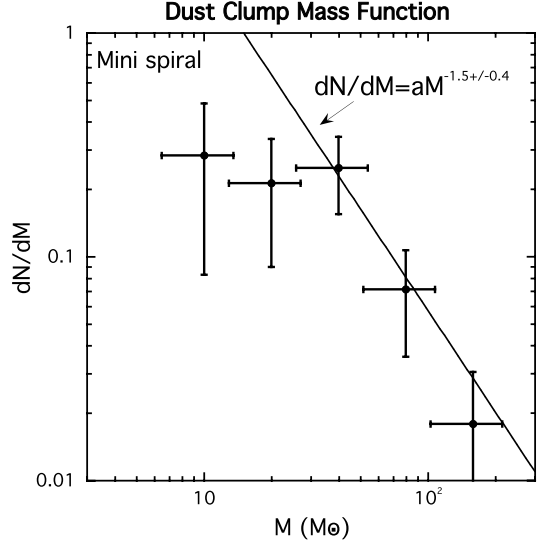}
 \end{center}
 \caption{Dust clump mass function (CMF) of the Mini-spiral at 340 GHz. 
The solid line shows the best-fit power-law function for massive three data points, $dN/dM\propto M^{-1.5\pm0.4}$.   }\label{Fig4}
\end{figure}
\begin{table}
 \tbl{Dust Clumps in the ``Mini-spiral".}{%
\begin{tabular}{cccccc}
\hline
RA offset$^a$&DEC offset$^a$&$d^b$&$M_c$&$\bar{n}$&$n_{min}^c$ \\ 
($\arcsec$)&($\arcsec$)&(pc)&(M$_\odot$)&(cm$^{-3}$)&(cm$^{-3}$)\\
\hline
3.43&0.984&0.147&10.2&6.4e+06&1.6e+10\\
7.38&-3.02&0.329&12.5&3.0e+06&1.4e+09\\
-2.83&-5.52&0.256&15.9&	6.3e+06&3.0e+09\\
0.02&-7.52&0.310&19.3&1.9e+06&1.7e+09\\
2.88&-2.52&0.158&24.1&5.6e+06&1.3e+10\\
6.28&2.48	&0.278&28.5&6.5e+06&2.3e+09\\
3.43&-1.02&0.147&34.1&2.6e+06&1.6e+10\\
1.67&-2.02&0.108&35.2&8.6e+06&4.0e+10\\
0.02&-2.52&0.104&35.6&6.4e+06&4.5e+10\\
3.43&-5.02&0.250&37.7&5.3e+06&3.2e+09\\
5.74&-6.02&0.343&41.1&2.9e+06&1.2e+09\\
13.1&-7.02&0.612&45.1&7.9e+06&2.2e+08\\
7.38&4.98	&0.367&81.4&5.7e+06&1.0e+09\\
-6.90&0.48&0.285&88.8&4.2e+06&2.2e+09\\
5.74&3.99	&0.288&103.3&4.2e+06&2.1e+09\\
5.74&0.48	&0.237&105.1&4.3e+06&3.7e+09\\
-3.38	&-1.52&0.153&140.4&1.3e+07&1.4e+10\\
-4.04&-3.52&0.221&173.8	&1.1e+07&4.6e+09\\
       \hline
\end{tabular}}
\label{tab:first}
\begin{tabnote}
$^a$ Angular offset from Sgr A$^{\ast}$ of the center of the detected clump. 
$^b$ Projected angular distance from Sgr A$^{\ast}$.
$^c$ Minimun number density required to sustain against the tidal shear; $n_{min}=5\times10^7d^{-3}$.
\end{tabnote}
\end{table}

As the next step of the quantitative analysis of the dust emission, we search dust clumps in the 340 GHz continuum map using the {\it clumpfind} algorithm (\cite{Williams1994}).  We detected 18 clumps within the FOV at 340 GHz. 
To estimate the clump mass, $M_c$, from the integrated intensity at 340 GHz, $F_{340}$, we use the formula for optically thin thermal emission from dust, $M_c=F_\nu d^2/\kappa_\nu B_\nu(T_d)$, where $\kappa_\nu$ is the dust mass absorption coefficient, $\kappa_\nu=0.1\times(\nu/1200 GHz)^{\beta}$ (\cite{Hildebrand}).
We assume here the dust temperature of $T_d = 20$ K as a conservative value,  $\beta=2$, and the metallicity of $Z/Z_0 = 1$.
The most massive core has $\sim170$ M$_\odot$.  When we get a new dust temperature of the clump, $T_d^{\prime}$, the clump mass changes to $(B_\nu(20)/B_\nu(T_d^{\prime}))M_c\sim(20/T_d^{\prime})M_c$ because of $h\nu/kT < 1$.
 Assuming a spherical dust particle, we estimate the mean number density using  $\bar{n}=M_c/\mu m_H/\frac{4\pi}{3}r_c^3$, where $\mu$ is the mean molecular weight, $\mu=2.3$.
The physical parameters of the clumps are shown in table 1.
Fig.4 shows the dust clump mass function (CMF).  
The function on the high-mass side seems to obey a power-law, which is usually used to depict the CMFs in the disk region of the Galaxy. The solid line in the figure shows the best-fit function for massive three data points, $dN/dM\propto M^{-1.5\pm0.4}$. 
Recent IR observations suggested that the stars in the vicinity of Sgr A$^{\ast}$ have continuously formed with a top-heavy IMF (e.g. \cite{Manes} ).
Although the slope of the Mini-spiral CMF on the high-mass side is steeper than that of the expected top-heavy IMF of the Central cluster, the steeper CMF can change into the flat IMF by the tidal shear of Sgr A*.
The mean gas number densities of the dust clumps are fairly less than the minimum gas density required for stability against the tidal shear  of the Sgr A$^{\ast}$ (\cite{Tsuboi2015}).  
The protostars already formed in the cloud could grow to stars finally because the typical number density of the protostars is larger than  the minimum gas density, on the other hand the remaining cloud gas will be ruined by the tidal shear.  
Consequently the IMF should become flatter than the CMF because the tidal destruction is more effective for lower mass cores.  

Based on the observed  tight correlation between the dust emission peaks and the OB/WR stars in the Northern-arm and the properties of the detected dust clumps, we would propose the following scenario as the origin of the Central cluster;
The stars including massive stars have formed in a molecular cloud located at a relatively large distance from Sgr A$^{\ast}$.   The massive star formation could be triggered by cloud-cloud collision because it can easily make high-mass stars (e.g. \cite{Tsuboi2015}). In addition, the cloud has considerably lost its angular momentum owing to the head-on collision. Consequently, the star-forming cloud has fallen into the immediate vicinity of Sgr A$^{\ast}$ with the time scale of $\sim10^5$ yr, comparable to the expected growth time of massive stars by accretion (e.g. \cite{Krumholz})
Now the falling cloud is being disrupted by the tidal shear of Sgr A$^{\ast}$ and the molecular gas in the cloud is ionized by UV continuum from the OB stars of the Central cluster. This is identified as the Mini-spiral. Because the period of the circular orbit around Sgr A$^{\ast}$ is $0.2-2\times10^4$ yrs at $R=0.1-0.5$ pc (see Table 2), the shearing motion would wind the Mini-spiral around Sgr A$^{\ast}$ closely within $\sim10^4$ yrs. 
Although  massive stars evolve quickly, low mass stars should be still on the stage of protostar. They may be the sources with a bow-shock appearance observed by JVLA (\cite{Yusef-Zadeh2015a}). 
A part of the massive stars formed in the cloud may be captured in the vicinity of Sgr A* and participate in the Central cluster. 
 The captured stars probably form a disk-like structure (Cf. \cite{Saitoh}). This is consistent with the observed disk-like distributions of the Central cluster (e.g. \cite{Paumard} ).

\begin{ack}  
We would like to acknowledge Dr. S. Nishiyama at Miyagi University of Education for useful discussion. The National Radio Astronomy Observatory is a facility of the National Science Foundation operated under cooperative agreement by Associated Universities, Inc. USA. This paper makes use of the following ALMA data:
ADS/JAO.ALMA\#2011.0.00887.S. ALMA is a partnership of ESO (representing
its member states), NSF (USA) and NINS (Japan), together with NRC
(Canada), NSC and ASIAA (Taiwan), and KASI (Republic of Korea), in
cooperation with the Republic of Chile. The Joint ALMA Observatory is
operated by ESO, AUI/NRAO and NAOJ. 
\end{ack}

\end{document}